%% file: dopstripe.tex
\documentstyle[prl,aps,multicol,epsf]{revtex}

\def\vc#1{{\bf #1}}
\newcommand{\be}{\begin{equation}}
\newcommand{\ee}{\end{equation}}
\newcommand{\bea}{\begin{eqnarray}}
\newcommand{\eea}{\end{eqnarray}}
\newcommand{\n}{\nonumber \\}
\newcommand{\tjm}{$t$-$J$ }
\newcommand{\htc}{high-T$_c$ }

\newcommand{\teps}{{\cal T}}  
\newcommand{\tteps}{\epsilon} 
\newcommand{\eps}{\varepsilon} 
\begin{document}
\draft

\title{Evolution of the stripe phase as a function of doping  \\
from a theoretical  analysis of angle-resolved photoemission data}

\author{Marc G.~Zacher, Robert Eder, Enrico Arrigoni, and
Werner Hanke}
\address{Institute for Theoretical Physics, University of  W\"urzburg,
97074 W\"urzburg, Germany}

\maketitle

\begin{abstract}
By comparing single-particle spectral functions of \tjm and Hubbard models
with recent angle-resolved photoemission (ARPES)
results for LSCO and Nd-LSCO, we can decide 
where holes go as a function of doping, and more specifically,
which type of stripe (bond-,
site-centered) is present in these materials 
at a given doping. For dopings greater than
about 12\% our calculation shows furthermore that the holes prefer to
proliferate out of the metallic stripes into the neighboring antiferromagnetic
domains.  
The spectra were calculated by a cluster perturbation technique, for 
which we
present an alternative formulation.
Implications for the theory for \htc
superconductivity are discussed.
\end{abstract}

\begin{multicols}{2}

\section{Introduction}
At present, stripes are at the heart of the debate concerning
the mechanism of superconductivity in high-temperature superconductors (HTSC).
There is clear experimental evidence for {\it static} stripes in Nd-doped 
LSCO (Nd-LSCO, for example La$_{1.48}$Nd$_{0.4}$Sr$_{0.12}$CuO$_4$)
from {\it elastic} neutron scattering experiments \cite{tranquada}.
The existence of a {\it dynamical} stripe phase in the ``real'' \htc compound
LSCO or even in YBCO has been conjectured from similar diffraction
patterns in the {\it inelastic} neutron scattering results
\cite{mook,dynlsco}. However, it has not been decided so far,
 whether the observed
pattern is due to one-dimensional spin-inhomogeneities (i.e. stripes)
\cite{mook} 
or  to two-dimensional incommensurable spin-waves \cite{keimer}. 
In the case of YBCO near optimal doping,
it has
been suggested that the incommensurable fourfold neutron scattering peak 
is 
due to the dispersion of the famous 41meV commensurable (found below
T$_c$ at momentum $\vc{k}=(\pi,\pi)$)
neutron scattering peak
to lower excitation energies around  $\vc{k}=(\pi,\pi)$ \cite{keimer}. 
From the theoretical point of view,
several numerical analyses, ranging from
Hartree-Fock~\cite{za.gu.89,ic.ma.99},
DMRG~\cite{white1} to dynamical mean-field theory~\cite{fl.li.00} indicate
that stripes can be produced by purely strong-correlation effects.
On the other hand, 
structural transformations~\cite{tranquada,bu.br.94,go.ro.99} 
as well as long-range Coulomb interactions~\cite{ki.em.96,ari}
may also play an important role in the formation of stripes.

In this article, we provide  numerical arguments showing that
an essential link in the chain of evidence for stripes is
 provided by angle-resolved photoemission
spectroscopy (ARPES): ARPES spectra show hardly any difference between 
LSCO ({\it dynamical} stripes candidate)
and Nd-LSCO ({\it static} stripe system) \cite{ndlsco,zhoudual,ino}.
This fact was first pointed out by a semiphenomenological argument by
Salkola et al.~\cite{sa.em.96}.
In a previous paper, we have
 shown that salient spectral features
of Nd-LSCO and LSCO can be explained by a model with static stripes
\cite{zacher}. Here, we will show that the spectra of LSCO and Nd-LSCO can be
almost quantitatively described by different types of stripe states
(i.e. site-centered, bond-centered)
for a wide variety of dopings. Since the experimental ARPES
spectra for LSCO and Nd-LSCO are so similar and their spectra can be 
quantitatively described
by stripe models, we argue that there must be stripes present in LSCO as well,
 at least in the underdoped to optimally doped regime.

If stripes are present as low-energy excitations 
in the \htc compounds,
they must affect the microscopic description of the superconducting
state, independently on whether they are an obstacle against superconductivity or even 
its driving force.
In this context,
 an important question is, whether stripes are bond-centered
or site-centered since, theoretically, bond-centered stripes have been shown 
to enhance superconducting pairing correlations \cite{em.ki.97.sg,ari}.
By analyzing the different scenarios with our technique and comparing the results with
ARPES spectra for different dopings,
 we can decide which type of stripe (bond- or
site-centered) 
better describes the spectrum as a function of doping.

Very recently, Zhou {\it et al.} \cite{zhoudual}
succeeded in measuring ARPES data on Nd-LSCO and LSCO systems with
different doping levels. In the doping region beyond 12\% the ARPES
results reveal a dual nature of the electronic structure: The straight 
segments 
forming  the ``Fermi surface'' in
energy-integrated photoemission and the $(\pi,0)$ low-energy
excitations, which have been
attributed to site-centered stripes for the 12\% doping case 
\cite{zacher}, are still present but overlaid with more two-dimensional
(2D) features reminiscent of a simple tight-binding band-structure for homogeneous
2D systems. Zhou {\it et al.} address the experimental question, whether the
two features originate from the mixing of two different phases or whether they
are intrinsic properties of the same stripe phase.
In the first case (phase separation) there should a non-stripe
phase with 
a very high carrier concentration (as estimated from the area inside the
Fermi surface) coexisting with
a phase of site-centered stripes. The authors
 remark that there is hardly evidence
for such a phase in Nd-LSCO and LSCO at the doping levels under consideration
\cite{ino,inopapers}. Another scenario consistent with the experimental
observation would be that the system forms more and more bond-centered
stripes upon increasing the doping
beyond 12\%. The possibility for the latter scenario 
was already indicated by our earlier calculations 
\cite{zacher}, where the spectrum of bond-centered stripes resembles
the diamond-shaped two-dimensional feature observed by Zhou {\it et al.}.
In the present article we will show that the evolution of the experimental
ARPES  with increasing
doping can be described by assuming that more and more 
bond-centered stripes are formed at the expense of site-centered ones
(as conjectured previously by Zhou {\it et al.}).

Our approach adopts the Cluster-Perturbation-Theory
(CPT) developed by Senechal {\em et al.}~\cite{senechal},  
which consists in splitting the infinite lattice into clusters which
are treated by exact diagonalization. The inter-cluster hopping terms
are then treated perturbatively, so that one eventually approximates the
infinite lattice. 
This method takes into account exactly the local correlations, 
which probably are the most important ones in these systems, and at the same
time makes all $k$ points of the Brillouin Zone available.
In addition, this is an ideal method to deal with a ``larger unit
cell'' such as the one present in the  stripe phase.
One should, however, mention that, at this level,  the
 method is not  appropriate 
 to explore stripe
stability for a given model. 
As a matter of fact, this is not the aim of the present work.
Rather, via this CPT approach we 
enforce a stripe pattern by connecting clusters with different hole
dopings [see Fig.~(\ref{figconf})], and study its spectral function to 
compare with ARPES experiments. A similar approach has been taken in 
Ref. \cite{piotr}, where the hole spectral function for site-centered
stripe patterns was calculated within the string picture.

Our paper is organized as follows: In section~\ref{tecn}, we 
explain in detail how the CPT 
is applied to the stripe phase. In section~\ref{nume}
 our numerical results are presented and
compared with ARPES spectra. The content of this article is 
summarized in section~\ref{conc}. 
Finally, the appendix presents 
an alternative derivation 
of the CPT equation, by means of a mapping onto a hard-core fermion
model.
In this appendix, we show that the CPT approximation amounts to
neglecting two-particle excitations within this model.

\section{Technique}
\label{tecn}

The computational technique for our calculation of the 
single-particle spectral weight $A(\vc{k}, \omega)$ and for the 
Green's function is a special application of the 
CPT for inhomogeneous systems.
This method is based on
a strong-coupling perturbation expansion of the (Hubbard model's)
one-body hopping operators linking the individual unit-cells \cite{senechal}.
At lowest order in this expansion, 
the Green's function of the
infinite lattice can be expressed (in matrix form) as: 
\begin{equation}
G^\infty (\vc{P},z) = \frac{G^{cluster}(z)}{1 - 
\tteps(\vc{P}) G^{cluster}(z)} \; ,
\label{eqcpa} 
\end{equation}
where the matrices $G^\infty$, $G^{cluster}$, and $\tteps(\vc{P})$ will be
defined around Eq.~(\ref{eps}).
The authors of ref.\cite{senechal} have pointed out that the above 
formula becomes exact in the limit of vanishing interaction ($U/t = 0$)
and, obviously, in the atomic limit ($t = 0$), and can thus be considered as an
interpolation scheme between $t\rightarrow \infty$ and $t\rightarrow 0$.
Notice, however, that this formula {\it does not} become exact in the
$U \to \infty$ limit.
In the interesting regime where the
interaction and the hopping are of the same order of magnitude S\'en\'echal 
{\it et al.}
have shown numerically
that Eq.(\ref{eqcpa}) gives as an accurate interpolation
between the two limiting cases. Additionally, we have shown in our previous
article \cite{zacher} that the cluster perturbation technique is
ideally 
suited to study 
inhomogeneous systems such as the stripe state in the \htc compounds.

\begin{figure}
\epsfxsize=8cm
\epsffile{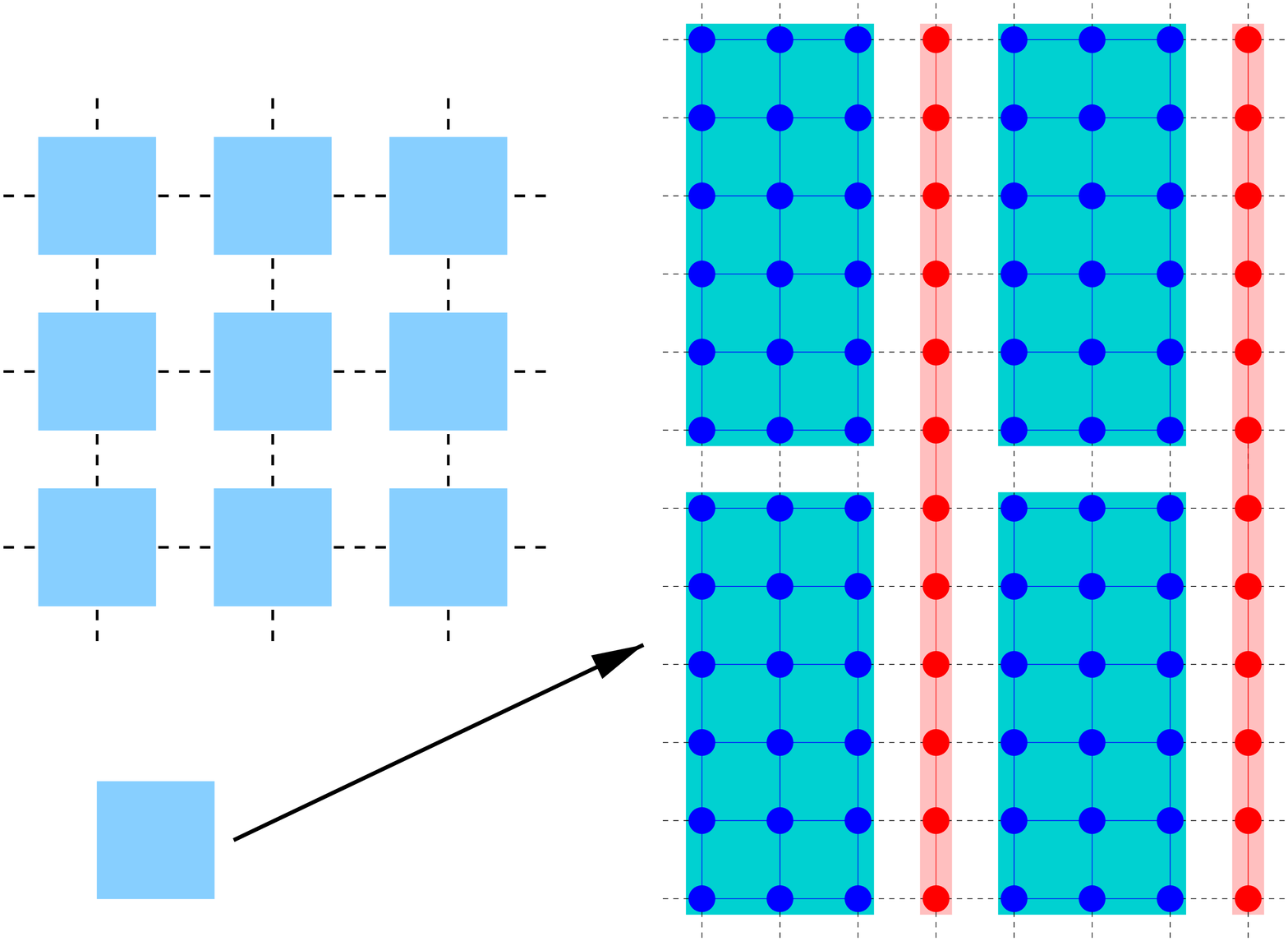}
\caption{Visualization of the cluster perturbation approach for stripes:
The infinite lattice is divided into unit cells. The unit cells differ for different
stripe configurations. 
As shown here, the unit cell for a 3+1 configuration consists of two $3 \times 6$ ladders on
top of each other next to a 12-site chain. The 3-leg ladders on the left half of the
unit-cell were diagonalized with a staggered magnetic field which was oriented in the
opposite direction of the one on the right half.
The hopping terms connecting the exactly--solved clusters ($3\times 6$ and $1 \times 12$) 
as well as the hoppings connecting the unit cells are included via the cluster perturbation
technique.
}
\label{figcpt}
\end{figure}

In order to deal with stripes,
the infinite lattice is divided into unit cells of equal size
(Fig.(\ref{figcpt}a)). 
The unit cell is further divided into independent blocks to incorporate  the
stripe topology:
In the example of Fig.(\ref{figcpt}), we are interested in a site-centered 
(``3+1'') configuration with quarter-filled metallic chains alternating with 
half-filled 
3-leg ladders.
The 3-leg ladders (here $3 \times 6$ blocks) 
and quarter-filled chains (here $1
\times 12$ with 6 holes, see Fig.(\ref{figcpt}b))
are solved by exact diagonalization, yielding the 
single-particle
spectral function of the block.
The individual Green's functions of the blocks 
forming a unit-cell cluster
are combined to form the
 Green's function of the unit cell $G^{cluster}$ at  ``order zero'',
 i. e., in which 
the intra-cluster hopping terms are set to zero.
In a second step,
the intra-cluster hopping connecting the individual blocks both within
the same unit cell and in different unit cells
(dashed
lines in Fig.(\ref{figcpt}b)) 
are incorporated via the cluster
perturbation technique.
 This yields the desired Green's function of 
the infinite lattice.
Where it was technically feasible,
we doubled the unit cell (as in Fig.(\ref{figcpt})) 
and diagonalized
two 3-leg ladders with a staggered magnetic field pointing in
opposite directions, resulting in a $\pi$-phase shifted
(between the AF domains)
N\'eel order in the final configuration.
This site-centered ``$3 + 1$'' configuration with  
$\pi$-phase shifted N\'eel order 
of stripes was first suggested
by Tranquada {\it et al.} \cite{tranquada}.
Bond-centered stripes, on the other hand, are modeled by
2-leg ladders with alternating filling. 
In the following, we will refer to this
bond-centered configuration as ``$2 + 2$''.

Holes can propagate out of the metallic stripes into 
the AF insulating domains via the inter- and intra-unit-cell hoppings
(dashed lines in Fig.(\ref{figcpt})). 
As explained above, our method consists in ``forcing'' the stripe structure
``by hand'' in order to study the effects  of this structure on the
photoemission spectrum. 
The different hole concentration in the ``metallic'' and in the
``antiferromagnetic'' region is achieved by adjusting independently
the chemical potential of the individual blocks so that the desired
hole density (see Fig.~\ref{figconf}) is obtained.
 This corresponds to introducing an on-site energy shift
between the two blocks
$\Delta  \approx 1.5 t$. Physically, this shift corresponds to the
energy of the stripe formation, possibly produced
by a combined effect of strong correlation and of 
 lattice distortion occurring in the low-temperature
tetragonal (LTT) phase~\cite{tranquada}.
Of course, without such an energy shift it would be impossible to
obtain such large density oscillations, such as the ones shown in
Fig.~\ref{figconf}. Indeed, in a homogeneous system the amplitude of charge
oscillations remains of the order or less than $0.1$, as shown, e. g., 
by DMRG calculations\cite{white1,ari}.

Alternatively equation(\ref{eqcpa}) can be obtained by expressing the fermionic
creation/annihilation operator $c^{(\dagger)}_i$ in terms of fermionic 
creation and
annihilation operators $d^{(\dagger)}_\alpha$ representing the 
photoemission and inverse photoemission target states $\vert \alpha
\rangle$ of the diagonalized cluster. This derivation of Eq. (\ref{eqcpa})
is presented in the appendix.

In Eq. (\ref{eqcpa}), $\vc{P}$ is a superlattice wave vector and $G^\infty$ 
is the Green's function of the ``$\infty$-size'' 2D system, 
however, still in a hybrid representation: real space 
within a cluster and Fourier-space between the clusters. 
This is related to the fact that $G^\infty(\vc{P}, z)$ is now an 
$M\times M$ matrix in the space of site indices (in the 
inhomogeneous stripe configuration of Fig.(\ref{figcpt}b) $M 
= 2 \times (3 \times 6) + 1 \times 12 = 96$). Likewise, $\tteps(\vc{P}
)$ and $G^{cluster}$ are $M\times M$ matrices in real 
space with $\tteps(\vc{P})$ standing for the perturbation. 
For the situation in Fig.(\ref{figcpt}b), the only nonzero elements of
the hermitian matrix
$\tteps(\vc{P})$ are  represented by the dashed bonds
in the figure:
\be
  \tteps(\vc{P})_{l,m} = \left\{ \begin{array}{r@{\quad:\quad}l}
  -t & \mbox{dashed bonds inside cell} \\
  -t e^{\pm i P_{x/y}} & \mbox{bonds connecting cells}\\
  0 & \mbox{elsewhere}
  \end{array} \right. \;.
\label{eps}
\ee
In order to facilitate  diagonalization of the individual clusters,
we used 
 periodic boundary conditions along
the stripe direction. 
In principle, this introduces hopping terms, which are not present in the 
infinite lattice. However, this is not a problem, since these terms
can be consistently removed  
 perturbatively by subtracting corresponding terms to the 
 matrix elements of $\tteps(\vc{P})$.

A complete Fourier 
representation of $G^\infty$ in terms of the original 
reciprocal lattice 
then yields the cluster 
perturbation theory (CPT) 
approximation~\cite{senechal}.

To allow for larger block sizes, we actually diagonalized 
the \tjm model on the blocks to obtain the (block-) Green's functions
as an approximation for the Hubbard model's (block-) Green's function.
The \tjm Hamiltonian is defined as
\be
H = - t \sum_{\langle i,j\rangle,\sigma} (\hat c^\dagger_{i,\sigma}
\hat c^{}_{j,\sigma}  + \mbox{h.c.}) + J
\sum_{\langle i,j\rangle} ( \vc{S}_i \vc{S}_j - \frac{n_i n_j}{4} ).
\ee
The sums run over
all nearest neighbor pairs $\langle i,j\rangle$.
No double occupancy is allowed.
We have chosen the commonly accepted values
$J/t = 0.4$ and $t \approx 0.5 eV$.

\begin{figure}
\epsfxsize=8cm
\epsffile{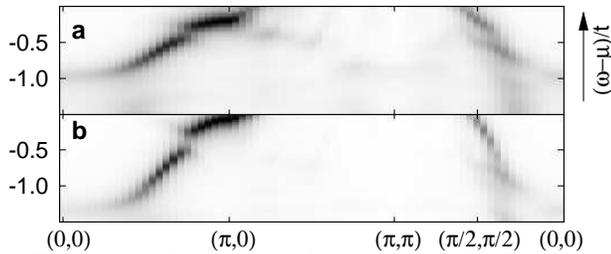}
\caption{
Single particle spectral function for a 2+2 bond-centered stripe
configuration at 12\% doping: 
Comparison of 
CPT calculations based on exact
diagonalizations of two 2-leg \tjm ladders with $J= 0.4 t$ (a) and 
on exact
diagonalizations of two 2-leg Hubbard ladders with $U= 8 t$ (b). 
The gray scale represents the weight of spectral function at the specific
($k, \omega$)-point with dark areas
corresponding to high spectral weight.
}
\label{figdenstjhubb}
\end{figure}

\begin{figure}
\epsfxsize=8cm
\epsffile{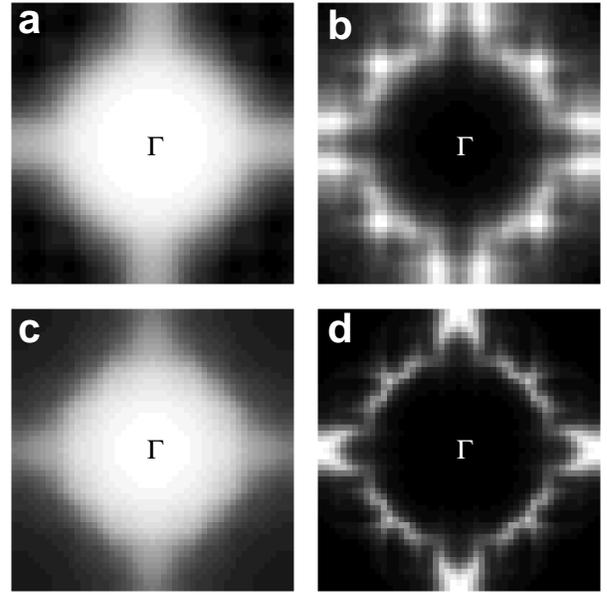}
\caption{Integrated spectral weight of bond-centered stripe configurations
at 12\% doping; comparison of CPT calculations based on exact
diagonalizations of \tjm ladders (a,b; $J = 0.4 t$) and Hubbard ladders
(c,d; $U = 8 t$):
(a,c) total integrated weight in photoemission ($n(\vc{k})$), 
(b,d) low energy excitations 
(integrated weight in $E_F-0.2 t < \omega < E_F$).
The data are plotted for the whole Brillouin zone with the $\Gamma$-point in
the center. The result of the stripe calculations have been symmetrized to
account 
for the differently oriented stripe domains in real materials. Regions of
high spectral
weight correspond to white areas. 
}
\label{figfermitjhubb}
\end{figure}

The quality of this approximation is tested by comparing single-particle
spectral functions of 2+2 bond-centered
stripe configurations at 12\% doping based on diagonalizations of both
Hubbard- and \tjm models. In Fig.(\ref{figdenstjhubb}),
 we show $A(\vec
k,\omega)$ for the standard walk through the Brillouin zone. One observes
that the result for the Hubbard model ($U= 8 t$, Fig.
(\ref{figdenstjhubb}a)) is very similar to the \tjm model result ($J =0.4
t$, Fig.(\ref{figdenstjhubb}b)): The dispersion is two-dimensional,
metallic-like and comparable to a tight-binding dispersion. The only 
difference between the figures is quantitative,
namely, the different Fermi velocities and bandwidths. 
This difference is  predominantly due to the omission of the conditional
hopping terms $\sim J$\cite{ha.la.67} in the \tjm Hamiltonian.
Another reason may be 
due to the fact that the basic parameters of the models, the interaction
strengths $J$ and $U$ have not been fine-tuned to match each other.
The integrated spectral weight obtained from the two models is depicted in
Fig.(\ref{figfermitjhubb}). In both cases, the occupation in momentum
space $n(\vec k)$ is distributed almost isotropically around the 
$\Gamma$ point. The low-energy excitations for both models
indicate a two-dimensional LDA-like Fermi surface, however with 
additional weight at the $(\pm \pi,0)$, $(0,\pm \pi)$ points.
The additional $(\pi,0)$ features are sharper in the Hubbard-model case.

Turning on the perturbation allows the holes to travel between the blocks
and unit-cells. 
In Fig.(\ref{figrealsp}), we show the electron concentration (averaged in
the direction along the stripes) in the direction perpendicular to the
stripes before (thick lines) and after (bars) applying CPT for the stripe
configurations that are discussed in this article.
As can be seen,
the holes do not travel far from the domains that were 
originally defined, and 
the electron occupation hardly changes from the
unperturbed setup.
Therefore, the desired stripe configuration is conserved in our
approach.

\section{Numerical results}
\label{nume}

\begin{figure}
\epsfxsize=8cm
\epsffile{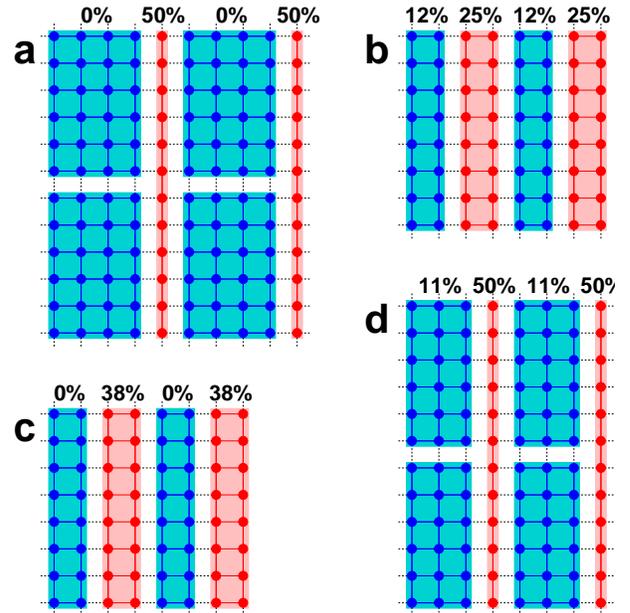}
\caption{Unit cells for different stripe configurations:
(a) 4+1 site-centered with 10 \% doping;
(b) 2+2 bond-centered with 19\% doping ($2 \times 8$
with 2 holes + $2 \times 8$ with 4 holes);
(c) 2+2 bond-centered with 19\% doping ($2 \times 8$
with 0 holes + $2 \times 8$ with 6 holes);
(d) 3+1 site-centered with 21 \% doping.
}
\label{figconf}
\end{figure}

\begin{figure}
\epsfxsize=7.5cm
\epsffile{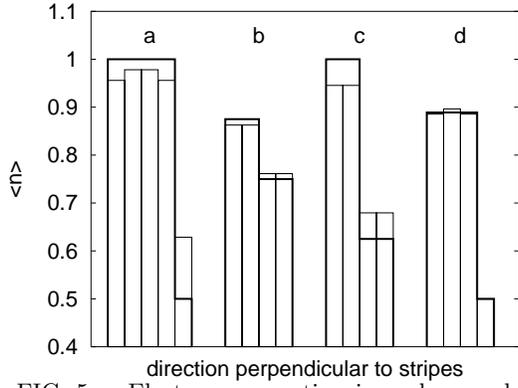}
\caption{
Electron occupation in real space before (thick lines) and
after (thin lines) application of CPT:
(a,b,c,d) according to {Fig.(\protect\ref{figconf})}.
}
\label{figrealsp}
\end{figure}

\begin{figure}
\epsfxsize=7.5cm
\epsffile{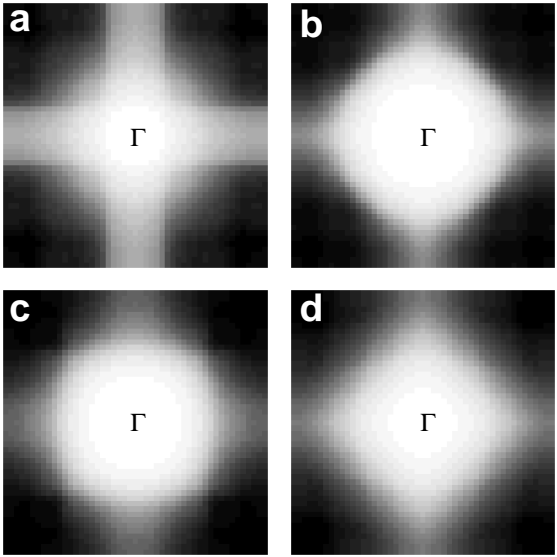}
\caption{
Total integrated spectral weight in photoemission $n(\vec k)$:
(a,b,c,d) according to {Fig.(\protect\ref{figconf})}.
}
\label{figdopnk}
\end{figure}

\begin{figure}
\epsfxsize=7.5cm
\epsffile{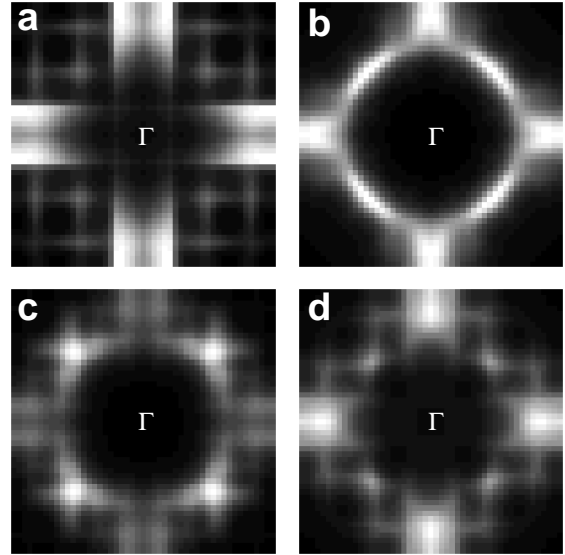}
\caption{
Integrated spectral weight around the Fermi surface:
(a,b,c,d) according to {Fig.(\protect\ref{figconf})}.
}
\label{figdopfermiw}
\end{figure}

\begin{figure}
\epsfxsize=7.5cm
\epsffile{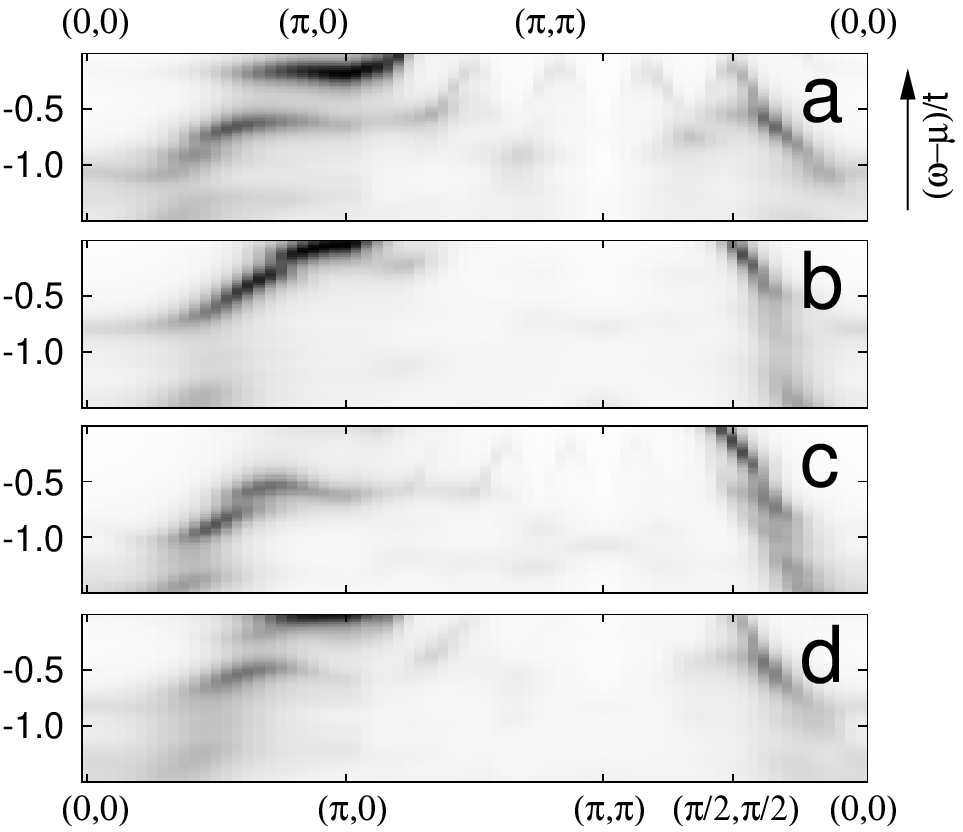}
\caption{
Density plot of the single particle spectral function:
(a,b,c,d) according to {Fig.(\protect\ref{figconf})}.
}
\label{figdopdens}
\end{figure}

We proceed to the discussion of the spectra for the underdoped region:
At 10 \% doping, the stripes have a charge periodicity of 5 lattice 
constants according to the Tranquada picture. This configuration has been
modeled by a $5 \times 12$ unit cell consisting of
two half-filled $4 \times 6$ systems (stacked on top of each other) and
a quarter-filled 12-site chain as displayed in Fig.(\ref{figconf}a).
Fig.(\ref{figdopnk}a) shows that the
spectral weight is confined in one-dimensional segments of the Brillouin
zone indicating a one-dimensional Fermi surface. The low-energy excitations
in Fig.(\ref{figdopfermiw}a)
are mainly located at the $(\pm \pi,0)$ and $(0,\pm \pi)$ points in
momentum space. In Fig.(\ref{figdopdens}a) the single-particle spectral
function for this doping is plotted directly and one observes the
characteristic stripe features that have been discussed in detail in
\cite{zacher}: a dispersionless band near $(\pi,0)$
crossing the Fermi surface at $(\pi,\pi/4)$ resulting from the
one-dimensional chain oriented in $y$-direction, the double-peak structure
at $(\pi,0)$ from the hybridization of the metallic band with the top of the
antiferromagnetic band and an excitation at $(\pi/2,\pi/2)$ at higher
binding energies than at $(\pi,0)$. The combined results for this doping
agree very well with the recent ARPES results by Zhou {\it et al.}
\cite{zhoudual}
giving further support for the static stripe picture by Tranquada:
at least below 12\% doping, charge carriers are only present in
quarter-filled chains that are alternating with half-filled
antiferromagnetic domains and an increasing of doping is realized by lowering
the distance between the chains and therefore reducing the effective size of the
antiferromagnetic domains.

The incommensuration of the quasi-elastic Neutron scattering does not
increase 
any more beyond a doping level of 12\%. 
The simple picture of one-dimensional chains moving closer together at the
expense of antiferromagnetic undoped domains thus cannot be valid in this regime.
For a doping of $1/8$, we have previously shown\cite{zacher} that the
ARPES data can be explained by assuming
that the static stripe system
(Nd-LSCO) is in a state of site-centered stripes,
where 3-leg antiferromagnetic ladders alternate with quarter-filled chains.
Here we will address the question how to describe the system for dopings
higher than $1/8$ or 12\%.

The ARPES results by Zhou {\it et al.} suggest
that LSCO and Nd-LSCO samples at 15 \% doping are in a state that is still
mainly in a 3+1 stripe phase since the ARPES results show all the features
that have been previously observed for the $x=0.12$ samples:
one-dimensional Fermi surface and low energy excitations located at
$(\pm \pi,0)$ and $(0,\pm \pi)$. In addition, however, 
low-energy excitations appear around the edge of a diamond located at
the center of the Brillouin zone. These excitations are connecting 
the $(\pi,0)$ features. Zhou  {\it et al.} conjecture that bond-centered
stripes are formed at the expense of site-centered ones, since it was shown
in our previous calculation that bond-centered stripes do indeed exhibit such
a diamond-shaped low-energy excitation pattern. 
This description is particularly interesting since bond-centered stripes in
contrast to site-centered ones have been shown to enhance superconducting
pairing correlations \cite{em.ki.97.sg}. 
Therefore, this picture may provide a link
to the doping dependence of the superconducting transition temperature of
LSCO.
However, to be able to relate bond-centered stripes to the diamond-shaped
low-energy excitations that appear in LSCO and Nd-LSCO at 15 \% doping,
one has to study bond-centered stripes at higher dopings than $x=15 \%$,
since, in a phase-separated state, they have to carry all
the additional holes that make up the overall doping of $x=15 \%$ of the
experimental sample. Here, we study three possibilities of ``overdoped''
stripe configurations as displayed in Fig.(\ref{figconf})
(labeled b,c,d consistent with the figure
label of Ref.~\cite{zhoudual}):

b) {\bf 2+2 bond-centered configuration (AF doped), Fig.(\ref{figconf}b)}:
Here, one ladder is at the filling
$\langle n \rangle = 0.75$ as in the case of 12\% doping and the other
ladder (previously undoped in the 12\% doping case) with a filling of
$\langle n \rangle = 0.875$ yielding an overall doping of 19 \%. In this
scenario 
{\it the doped region extends into the antiferromagnetic domain between
the charged stripes}. Technically, this configuration has been realized by
coupling a $2 \times 8$ ladder (2 holes) with a $2 \times 8$ ladder 
(4 holes).

c) {\bf 2+2 bond-centered configuration, Fig.(\ref{figconf}c)}:
Here, one ladder is at the filling
$\langle n \rangle = 0.625$ and the other
ladder stays half-filled as in the
12\% doping case, again yielding an overall doping of 19 \%. In this
scenario the excess holes further populate the charged stripe.
Technically, this configuration has been realized by
coupling a $2 \times 8$ ladder (no holes) with a $2 \times 8$ ladder 
(6 holes).

d) {\bf 3+1 site-centered configuration, Fig.(\ref{figconf}d)}:
Here, the chain is quarter-filled
($\langle n \rangle = 0.5$) as in the case of 12\% doping and the
three-leg ladder
(previously undoped in the 12\% doping case) has a filling of
$\langle n \rangle = 0.89$ yielding an overall doping of 21 \%. In this
scenario, the doped region extends into the antiferromagnetic domain between
the charged stripes (as in case b).
Technically, this configuration has been realized by
coupling two $3 \times 6$ ladders (2 holes each) with a $1 \times 12$ chain 
(6 holes).

In Fig.(\ref{figdopnk} b,c,d) the electron occupation in momentum space is
displayed. Neither of the three configuration shows the typical stripe
signatures (one-dimensional distribution of spectral weight) but the weight
is more or less isotropically distributed around the $\Gamma$-point. The
spectral weight in Fig.(\ref{figdopnk} c) is almost circularly distributed
and resembles $n(\vec k)$ of the free electron gas. The weight distribution
in Figs. (\ref{figdopnk} b,d), on the other hand, is more of a diamond shape.
In the ``phase-separated'' picture, we expect the
 spectral weight stemming from the above
configurations (b,c,d) to be superimposed onto the spectral weight coming
from the dominating 3+1 site-centered structure which is all concentrated
in the one-dimensional segments in momentum space. This might be the reason
that in the actual experiment Zhou and coworkers can only resolve the
latter. 

For the low-energy excitations, the situation is different since a much
smaller energy window is integrated and therefore the experiment is more
sensitive to smaller amounts of spectral weight: Here, the results of the
calculation are shown in  Figs. (\ref{figdopfermiw} b,c,d). The low-energy
excitations of the 2+2 configuration c are concentrated around the $(\pm
\pi/2,\pm \pi/2)$ points. Overlaying these excitations with the 3+1
site-centered features would not yield the experimentally observed diamond
structure connecting the $(\pi,0)$ excitations and therefore this
configuration can be discarded. In contrast, the other 2+2 configuration
b, where the extra holes are populating the antiferromagnetic domains
does indeed show the experimentally observed diamond shape. In this setup,
the $(\pi,0)$ features are present as well. They further enhance the
low-energy excitations in this region of momentum space which are due
to the domains that are still in the
3+1 site-centered configuration (present at 12\% doping). 
The doped 3+1 site-centered configuration d, where the holes extend into the
antiferromagnetic region also has its low-energy excitations
distributed around a diamond centered at the $\Gamma$ point (Fig.
(\ref{figdopfermiw} d). However, its
features are not as sharp as in Fig.(\ref{figdopfermiw} b). 
This configuration cannot be so easily discarded and might be present in the
actual material.

For completeness, we display the single-particle spectral weight 
of the three
``overdoped'' stripe configurations in Figs. (\ref{figdopdens} b,c,d):
Configurations b and d show a two-dimensional tight-binding-like
dispersion, the difference being the large amount of spectral weight that
is concentrated near the Fermi level at $(\pi,0)$ for the bond-centered
configuration b, which was also visible in the low-energy excitation plot
(Fig.(\ref{figdopfermiw} b)). The other bond-centered configuration c,
where all doped holes are concentrated in one ladder, only has a Fermi-level
crossing at $(\pi/2,\pi/2)$ (``hole-pockets'')
consistent with Fig.(\ref{figdopfermiw} c).

\section{conclusion}
\label{conc}

In this work, the cluster perturbation technique and its
application to the stripe phase of \htc materials and related
compounds has been 
studied in detail.
The CPT has been applied to Hubbard and \tjm systems with different dopings
and stripe configurations.  The comparison of our results 
with recent ARPES data suggest that, in the case of LSCO and Nd-LSCO,
stripes are present over a wide doping range. For dopings below 12\% the system
consists of site-centered stripes, whereas for higher dopings more and more 
bond-centered stripes are present at the expense of site-centered ones.
In the case of bond-centered stripes at dopings higher than 12\%,
 we have provided evidence 
that the excess holes prefer to proliferate out of the stripes into the 
AF domain.

\section*{acknowledgments}
\label{ackn}
This work owes much to fruitful discussions with 
Z.-X. Shen and X.J. Zhou.
The authors acknowledge 
financial support from
BMBF (05SB8WWA1) and DFG (HA1537/20-1). The calculations 
were carried out at the high-performance 
computing centers HLRS (Stuttgart) and LRZ (M\"unchen).

\input{appendix.tex}

%
%
\bibliography{biblio}
\bibliographystyle{prsty-macros}

\newpage
\section*{figures}
\narrowtext

\end{multicols}

\end{document}

%% file: appendix.tex

\section*{Appendix}
\label{appe}
\def\cmu{\mu}
\def\cnu{\nu}
\def\calpha{\alpha}
\def\cbeta{\beta}

The CPT expression for Green's function of the infinite lattice, Eq. (\ref{eqcpa}), 
can be obtained in an alternative way by mapping the fermionic
operators onto ``quasiparticle'' operators which create the 
ionization and affinity
states from the ground state\cite{dorneich,go.ri.94}.
We discuss here for simplicity the homogeneous case, whereby the lattice 
is divided into equal
clusters, 
although the extension to our case of different clusters is straightforward.
Consider now the ground state $\vert G\rangle$ of the cluster
Hamiltonian (say with $N$ particles) and the excited states
 $\vert \calpha\rangle$ with $N+1$ and $N-1$ particles
with corresponding excitation energies (including chemical potential)
$\eps_\calpha$.
We can think the
 $\vert \calpha \rangle$ as being created from the ground state via
a fermionic  creation operator
\be
\vert \calpha \rangle =
d^\dagger_\calpha \vert G \rangle \, ,
\ee
which must satisfy the hard-core constraint
\be
\sum_\calpha d^\dagger_\calpha d^{}_\calpha \leq 1 \;.
\ee 
If one neglects two-particle excitations, 
the original fermion  operators
$c^{\dagger}_{n}$ 
($n$ is the combined site and spin index within the
cluster) can be expressed in terms of the $d^{\dagger}_\calpha$ and $d_{\calpha}$ as
\be
 c^\dagger_n = \sum_\cmu T^*_{n,\cmu} d^\dagger_\cmu + \sum_\cnu
S^*_{n,\cnu} d_\cnu
\label{eqcrep}
\ee
with
\bea
\label{ts}
  T_{n,\cmu} &=& \langle G \vert c_n \vert \cmu \rangle \n
  S_{n,\cnu} &=& \langle \cnu \vert c_n \vert G \rangle \, .
\eea
Here and in the following, we are
using the
label $\cmu$ for $N+1$-particle (inverse photoemission) states, and $\cnu$ for
$N-1$-particle (photoemission) states. Other labels will not
distinguish between them.
In terms of the $c_n$, 
the inter-cluster hopping part of the Hamiltonian has the general form
\be
\hat \teps = \sum_{n,m} \sum_{a,b} \teps_{n,m}(a,b) c^\dagger_n(a) c^{}_m(b) \,
\ee
where $a,b$ label the individual clusters.
Neglecting the two-particle excitations and  the constraint, the
Hamiltonian of the infinite lattice becomes
\be
\label{h}
  \hat H = \hat H_c + \hat \teps \, ,
\ee
where $\hat H_c$ is the intra-cluster Hamiltonian
\be
  \hat H_c = \sum_a \sum_\calpha \eps_\calpha \ d^\dagger_\calpha(a) d^{}_\calpha(a)
   \, ,
\ee
and we have introduced a cluster index $a$ for the $d_{\calpha}$.
The Hamiltonian Eq.(\ref{h}) is now quadratic in the $d_{\calpha}$
operators and can be readily solved exactly by Fourier transformation
of the inter-cluster part $\hat \teps$ in the Cluster variables $a,b$.

We  now show  that the resulting Green's function for the $c_n$
operators is given by Eq.(\ref{eqcpa}). 
For convenience, we first carry out a particle-hole transformation on
 the  operators $d^{\dagger}_{\calpha}$, 
\bea
  p^\dagger_\cmu &=& d^\dagger_\cmu  \n
  p^\dagger_\cnu &=& d^{}_\cnu \; ,
\eea
such that the new operators $p^\dagger$
all create
particles.
Eq. (\ref{eqcrep}) 
simplifies to
\be
\label{cqp}
  c^\dagger_n = \sum_\calpha Q^*_{n,\calpha}\ p^\dagger_\calpha\, ,
\ee
where the matrix $Q_{n,\calpha}$ is given in terms of the $T$ and $S$
matrices in Eq.~(\ref{ts}) as
$Q_{n,\cmu}= T_{n,\cmu}$  and $Q_{n,\cnu}= S_{n,\cnu}$ .
The particle-hole transformation affects
$\hat H_c$ which becomes
\be
\hat H_c = \sum_\calpha \eps_\calpha \eta_\calpha \sum_a p^\dagger_\calpha(a)
 p^{}_\calpha(a) +\mbox{const.}\, ,
\ee
with
\bea
  \eta_\cmu &=& +1 \n
  \eta_\cnu &=& -1 \,,
\eea
and
 $\hat \teps$ is transformed to 
\be
  \hat \teps = \sum_{a,b} \sum_{n,m} \sum_{\calpha,\cbeta} 
      \teps_{n,m}(a,b) \ Q^*_{n,\calpha} Q^{}_{m,\cbeta} \ p^\dagger_\calpha(a)
      p^{}_\cbeta(b) \, .
\ee
The total Hamiltonian Eq.~(\ref{h}) can thus be written in the 
form 
\be
\label{hpp}
  \hat H = \sum_{a,b} \sum_{\calpha,\cbeta} p^\dagger_\calpha(a)
h_{\calpha,\cbeta}(a,b) p^{}_\cbeta(b)
\ee
with
\be
  h_{\calpha,\cbeta}(a,b) = \delta_{\calpha,\cbeta} \eps_\calpha \eta_\calpha
  + \left( Q^{\dagger} \teps(a,b) Q \right)_{\calpha,\cbeta} \, .
\ee
The Green's function for the $p$ is readily evaluated
from Eq.~(\ref{hpp})
\be
  \langle\langle p_\calpha(a), p_\cbeta^\dagger(b)\rangle\rangle
 = \left( z- h \right)^{-1}_{\calpha a,\cbeta b} \;,
\ee
where we have considered the terms within braces as matrices in the
indices $\calpha a,\cbeta b$, whereby the complex frequency $z$ is proportional to the
identity matrix.
The Green's function for the ``true'' particles $c$ is readily obtained by inserting the
transformation 
Eq.~(\ref{cqp})
\be
\label{ganbm}
 G^{\infty}_{a n,b m} \equiv \langle\langle c_n(a), c_m^\dagger(b)\rangle\rangle
 = \left( Q \left( z - h \right)^{-1} Q^{\dag} \right)_{\calpha a,\cbeta b}
  \, .
\ee
We now want to show that Eq.~(\ref{ganbm}) is equivalent to Eq.~(\ref{eqcpa}).

Consider first the {\em exact} Green's function of a single Cluster
\be
 G^{cluster}_{a n,b m} \equiv  \delta_{a,b} \langle\langle c_n(a),
 c_m^\dagger(a)\rangle\rangle_c \;.
\ee
By inserting Eq.~(\ref{ts}) in its Lehmann representation, one obtains
\bea
G^{cluster}_{a n,a m} &=& \left[T (z - \eps)^{-1} T^{\dag} + S (z +
  \eps)^{-1} S^{\dag}\right]_{n,m} \n
\label{gcq}
&=&
\left[Q (z - \eps \eta)^{-1} Q^{\dag} \right]_{n,m} \;,
\eea
where the matrices $\eps$ and $\eta$ are diagonal matrices in the indices
$\alpha,\beta$ with values $\eps_\alpha$, and $\eta_\alpha$,
respectively.

Using the anticommutation rules, it is now straightforward to show
that the product 
$Q \ Q^{\dag} $
  is equal to the identity matrix $I$.
Let us assume for a moment that 
 $Q$ is a square matrix,
 so that $Q$ is a unitary matrix, and $Q^{\dag} Q = I$ as well.
In this way, Eq.~(\ref{ganbm}) can be readily inverted yielding
 (we consider the matrix $Q$ as the identity matrix in
the $a,b$ indices, i. e. $Q_{n a,\beta b}=\delta_{a,b} Q_{n,\beta}$) 
\bea
\label{gm1}
 G^{\infty \, -1} 
 &=&  Q \left( z - h \right) Q^{\dag} \n
 &=&
Q (z - \eps \eta - Q^{\dag} \teps Q ) Q^{\dag} 
= G^{cluster \, -1} - \teps 
\eea
which is equivalent to Eq.~(\ref{eqcpa}), if one considers that
$\tteps(\vc{P}) $ (Eq.~\ref{eps})
  is nothing but the Fourier transform of $\teps$ in the cluster
 coordinates $a$ and $b$.

Unfortunately, $Q$ is in general not a square matrix, as there are
more single-particle excited states as particles. Therefore, one
cannot easily invert Eq.~(\ref{ganbm}).
Nevertheless, this problem can be readily overcome. We sketch the main 
procedure below.
The  matrix $Q_{n,\alpha}$, say with dimensions $R\times S$,  consists of $R$
row orthonormal vectors  $v^{(n)}_{\alpha}$. By appending the remaining $S-R$
orthonormal vectors to $Q$, one obtains a 
 square matrix $\bar Q$ which is now unitary. 
For the
 ``extended'' Green's functions $\bar G^{\infty}$ and $\bar G^{cluster}$,
obtained by replacing $Q\to \bar Q$ in Eqs.~(\ref{ganbm}) and
(\ref{gcq}), respectively, the relation 
Eq.~(\ref{gm1}) obviously holds.
 It is now a matter of matrix algebra to show that $G^{\infty}$ and $G^{cluster}$ are
 given by the ``upper left'' $R\times R$ blocks (in the $n,m$ indices) 
 of the respective ``extended matrices, i. e.
\be
G^{\infty}_{a n,b m } = \bar G^{\infty}_{a n,b m } \quad \mbox{for $n,m\leq R$} \;,
\ee
and the same for $G^{cluster}$.
The last line of 
Eq.~(\ref{gm1}), thus, holds for the case of non-square matrices $Q$
as well.

In summary, we have shown that the CPT is equivalent to a 
mapping onto a model of hard-core fermions describing single-particle 
excitations from the ground state of the cluster.  This fact suggest
an improvement of the method whereby two-particle processes, such as
spin, charge, or pair excitations are taken into account by
introducing appropriate 
hard-core bosons.
